\documentclass{article}

\usepackage{arxiv}

\usepackage[utf8]{inputenc} 
\usepackage[T1]{fontenc}    
\usepackage{hyperref}       
\usepackage{url}            
\usepackage{booktabs}       
\usepackage{amsfonts}       
\usepackage{nicefrac}       
\usepackage{microtype}      
\usepackage{lipsum}		
\usepackage{graphicx}
\usepackage{natbib}
\usepackage{float}
\usepackage{doi}
\usepackage{listings}
\usepackage{xcolor}
\definecolor{codegreen}{rgb}{0,0.6,0}
\definecolor{codegray}{rgb}{0.5,0.5,0.5}
\definecolor{codepurple}{rgb}{0,0,205}
\definecolor{backcolour}{rgb}{0.95,0.95,0.92}
\lstdefinestyle{mystyle}{
    backgroundcolor=\color{backcolour},   
    commentstyle=\color{codegreen},
    keywordstyle=\color{magenta},
    numberstyle=\tiny\color{codegray},
    stringstyle=\color{codepurple},
    basicstyle=\ttfamily\footnotesize,
    breakatwhitespace=false,         
    breaklines=true,                 
    captionpos=b,                    
    keepspaces=true,                 
    numbers=left,                    
    numbersep=5pt,                  
    showspaces=false,                
    showstringspaces=false,
    showtabs=false,                  
    tabsize=2
}
\title{Domain Analysis of Ethical, Social and Environmental Accounting Methods}
\author{ \href{https://orcid.org/0000}{\includegraphics[scale=0.06]{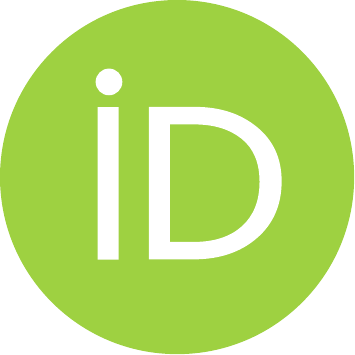}\hspace{1mm}Vijanti Ramautar}\\
	Department of Information and Computing Sciences\\
	Utrecht University\\
	Princetonplein 5, 3584 CC Utrecht \\
	\texttt{v.d.ramautar@uu.nl} \\
	\And
	\href{https://orcid.org/0000}{\includegraphics[scale=0.06]{orcid.pdf}\hspace{1mm}Sergio España} \\
	Department of Information and Computing Sciences\\
	Utrecht University\\
	Princetonplein 5, 3584 CC Utrecht \\
	\texttt{s.espana@uu.nl} \\
}

\renewcommand{\headeright}{Version 1}
\renewcommand{\undertitle}{Version 1}
\renewcommand{\shorttitle}{ESEA methods}

\hypersetup{
pdftitle={ESEA Methods},
pdfsubject={ICS},
pdfauthor={Vijanti, Ramautar, Sergio, España},
pdfkeywords={Sustainability, Modelling, Accounting},
}

\begin{document}
\maketitle

\begin{abstract}
Ethical, social and environmental accounting is the practice of assessing and reporting organisations' performance on environmental, social and governance topics. There are ample methods that describe how to perform such sustainability assessments. This report presents a domain analysis of ethical, social and environmental accounting methods. Our analysis contains 21 methods. Each method is modelled as a process deliverable diagram. The diagrams have been validated by experts in the methods. The diagrams lay the foundation for further analysis and software development. In this report, we touch upon the ethical, social and environmental accounting method ontology that has been created based on the domain analysis.
\end{abstract}

\section{Introduction}
Ethical, social and environmental accounting (ESEA) methods guide responsible entities in performing sustainability assessments and reporting on the assessment results. Typically organisations report on their performance regarding ethical, social and environmental dimensions. The contributions of this report include a detailed domain analysis of 21 ESEA methods. For each, method, we propose a process deliverable diagram (PDD) that depicts the method from two perspectives: the process part focuses on the activities of the method, and the deliverable part focuses on the products used as input or produced as output by the activities (thus constituting a conceptual data model for the method) \cite{weerd2009meta}.The diagrams are created as part of the Software for Organisation Responsibility research line. We base the models on method documentation and official information published by the entity that developed the method. The PDDs are validated with experts in the respective ESEA methods. After the validation interviews, we update and improve the PDDs if necessary. The validated PDDs are used to create activity and concept comparisons. For this, we adapt the method comparison approach \citep{weerd2007developing}. We provide a discussion of the results and present a metamodel that abstracts the 21 ESEA methods. We name this the openESEA metamodel since it acts as a domain ontology that serves as the foundation for engineering the openESEA domain-specific language (DSL). The openESEA metamodel and the DSL are presented in a technical report \citep{TR_modelling_language}. An earlier version of the metamodel and DSL are presented in \cite{espana2019model}.

The report structure is the following. Section \ref{sec:research_method} lists the research questions and provides an overview of the research method. Section~\ref{sec:PDDs} contains the domain analysis of ESEA methods, including all PDDs and brief descriptions of each of the ESEA methods. Section~\ref{sec:discussion} discusses the results of the domain analysis of ESEA methods. The result includes the openESEA metamodel. In Section~\ref{sec:conclusion} we present the conclusions of this research.

\section{Research method}
\label{sec:research_method}

\subsection{Research questions}
This report answers the following research questions.
\begin{itemize}
    \item[RQ1] What is the state of the art in ethical, social and environmental accounting methods?
    \item \begin{itemize}
        \item[RQ1.1] Which ESEA methods exist?
        \item[RQ1.2] What are the commonalities and differences in ESEA methods in terms of the process and data structure?
    \end{itemize}
    \item[RQ2] What are requirements for engineering a be part of a domain-specific modelling language for specifying ESEA methods?
    \item \begin{itemize}
        \item[RQ2.1] Which method constructs should be part of a domain ontology for ESEA methods?
        \end{itemize}
\end{itemize}

\subsection{Overview of the research method}
\begin{figure}[H]
     \centering
     \includegraphics[width=\textwidth]{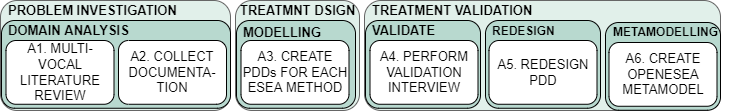}
     \caption{An overview of the research method}
     \label{fig:research_method}
 \end{figure}

 Figure~\ref{fig:research_method} provides an overview of the research method. We have performed a multi-vocal literature review (activity A1) to search for ESEA methods. We collected both scientific and grey literature because in this domain we find much valuable information coming from practitioners \cite{garousi2016need}. Practitioners can be part of networks of social enterprises such as B Corporations \cite{BIA} and institutions such as the United Nations \cite{williams2004global}. 
 
For each ESEA method identified in the literature, we have searched for and collected as much documentation as possible (activity A2). We are interested in sources of diverse nature, such as method manuals, standards related to the methods, websites providing instructions to users, tools supporting the methods, examples of the application of the methods (e.g. sustainability reports from companies applying the method).

Based on the collected documentation in activity A2 we have created a process deliverable diagram for each ESEA method (activity A3). The PDDs are validated with experts of the methods (activity A4). In the validation interview we present experts with the PDDs and ask them whether the process activities are complete and modelled in the correct order. Additionally, we ask them to validate whether all relevant deliverables are present in the PDD. After the validation interviews we improve and redesign the PDDs (activity A5) based on the expert feedback. Based on the validated PDDs we create the openESEA metamodel \cite{TR_modelling_language}.
 
\section{Domain analysis of ethical, social and environmental methods}
\label{sec:PDDs}
The following sections give a brief explanation of each of the analysed ESEA method. After the explanation the PDD of the corresponding method is depicted. For each PDD we have multiple versions, since ESEA methods evolve and so does our knowledge of said methods. In this report we present the latest versions of each of the method diagrams. Most diagrams are validated with experts on the method. However, not all diagrams are yet validated. Table~\ref{tab:Method_List} contains a list of all analysed ESEA methods.

\subsection{AA1000AS}
The AA1000 Assurance Standard (AA1000AS v3) is a method used by sustainability professionals for sustainability-related assurance engagements, to assess the nature and extent to which an organisation adheres to the AccountAbility Principles \cite{AA1000}. Initially we classified this method as an ESEA method. Currently we classify the method as an assurance standard, which covers the last part of the ESEA process.

\begin{figure}[H]
    \centering
    \includegraphics[width=0.8\textwidth]{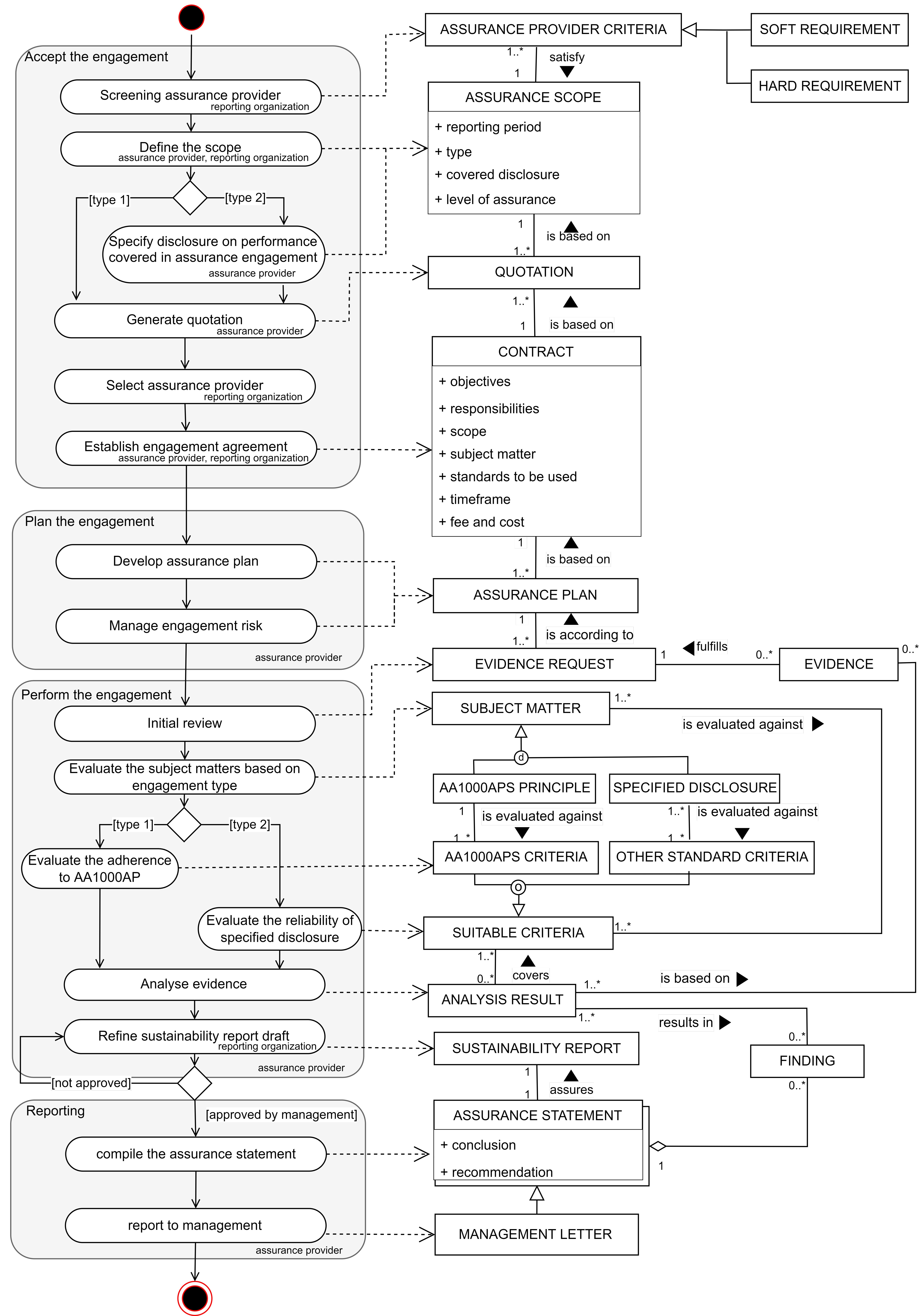}
    \caption{The PDD of AA1000AS}
    \label{fig:AA1000AS}
\end{figure}

\newpage

\subsection{B Impact Assessment}
The B Impact Assessment \cite{BIA} is a method developed by the not-for profit organisation B Lab. Application of the B Impact Assessment can result in an official B Corporation certification. In order to achieve the certification, organisation should meet the legal requirements and score at least 80 of the 200 points on the assessment. It is a private certification of for-profit companies, distinct from the legal designation as a Benefit corporation. B Corp certification is conferred by B Lab, a global nonprofit organisation.

\begin{figure}[H]
    \centering
    \includegraphics[width=0.9\textwidth]{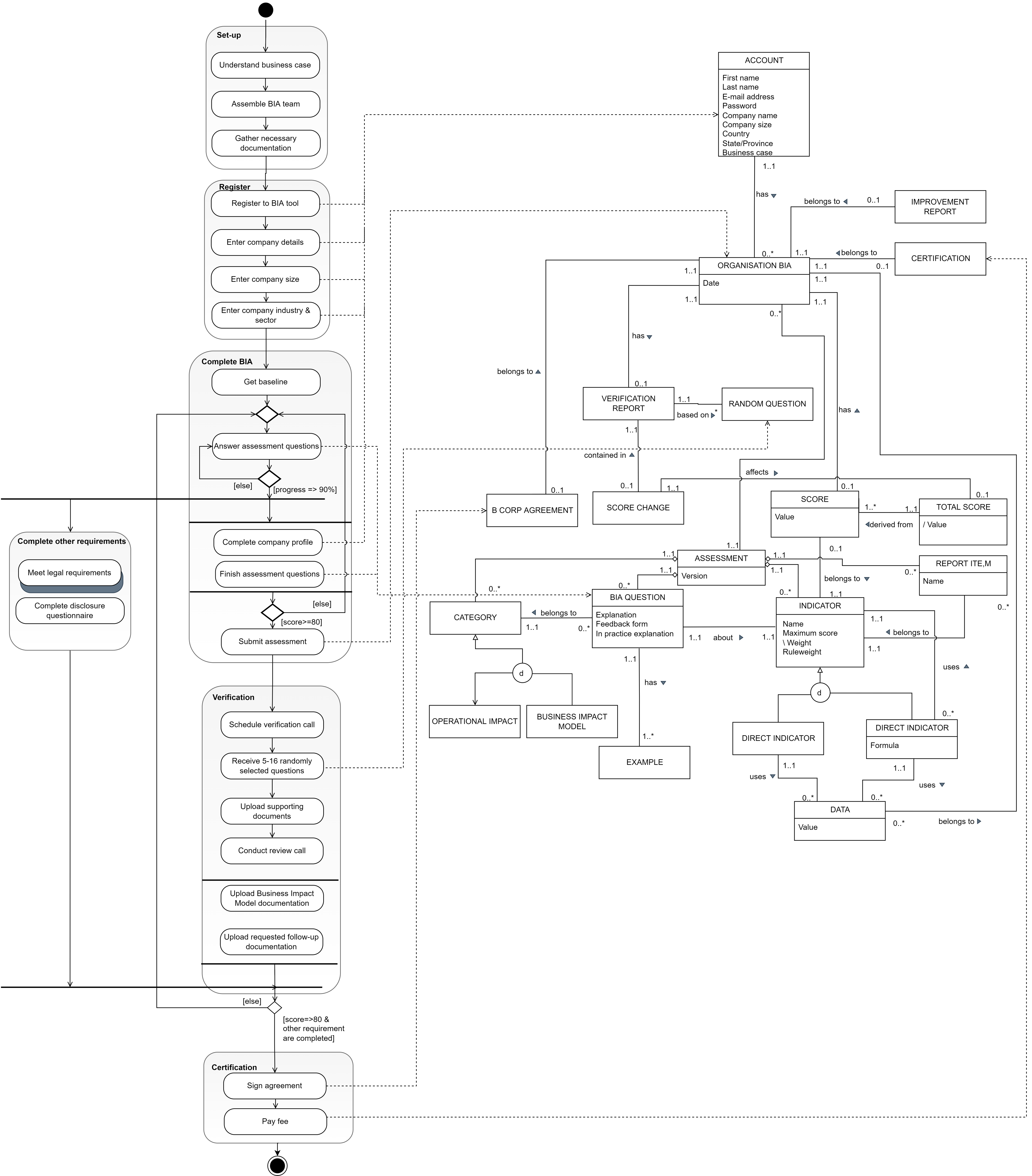}
    \caption{The PDD of the B Impact Assessment}
    \label{fig:BIA}
\end{figure}

\newpage

\subsection{CDP}
The CDP (formaly know as the Carbon Disclosure Project) is an international non-profit organisation based in the United Kingdom, Japan, India, China, Germany and the United States of America that helps companies and cities disclose their environmental impact \cite{CDP}. The PDD depicts the process of applying the CDP company programs. CDP (for companies) has three corporate questionnaires; climate change, forests and water security. The questionnaires provide a framework for companies to provide environmental information to their stakeholders covering governance and policy, risks and opportunity management, environmental targets and strategy and scenario analysis.       

\begin{figure}[H]
    \centering
    \includegraphics[width=\textwidth,height=0.8\textheight]{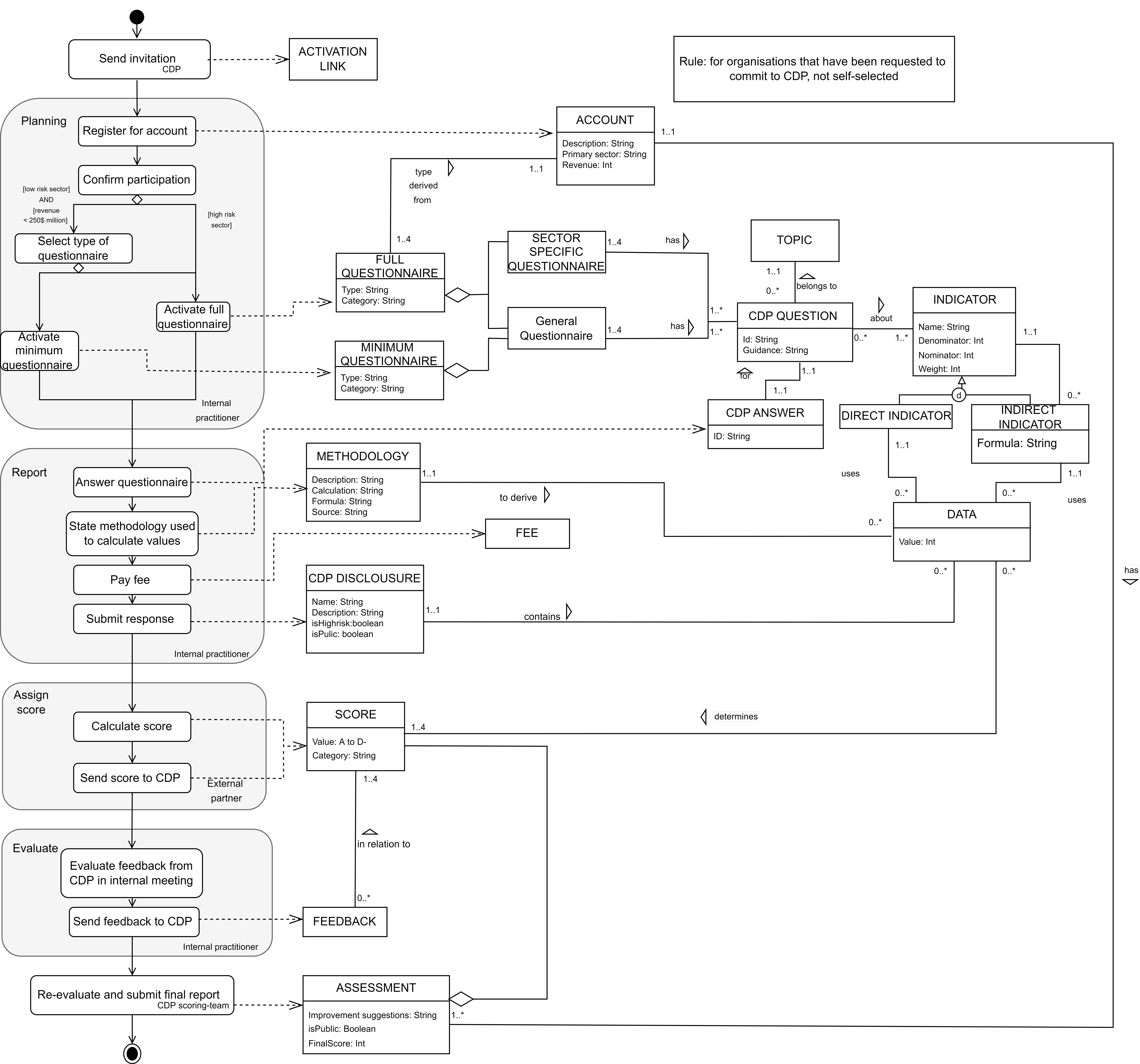}
    \caption{The PDD of the CDP method}
    \label{fig:CDP}
\end{figure}

\subsection{Common Good Balance Sheet}
The Common Good Matrix is a framework for the evaluation of business activities and an aid for organisational development \cite{felber2019common}. It describes 20 Common Good themes and gives guidance on how to evaluate based on Common Good principles. A Common Good Report is a comprehensive evaluation of a company‘s contribution to the common good, and is prepared as part of the reporting process. It should include a description of how the company‘s activities relate to each of the 20 common good themes. This will show how developed each value is within the company. Each theme describes how the individual values apply to the relevant stakeholder group. The Certificate documents an externally audited evaluation of the individual themes, gives an, and presents this in the layout of the Matrix. Together, the Common Good Report and Certificate represent the Common Good Balance Sheet.
 
\begin{figure}[H]
    \centering
    \includegraphics[width=\textwidth]{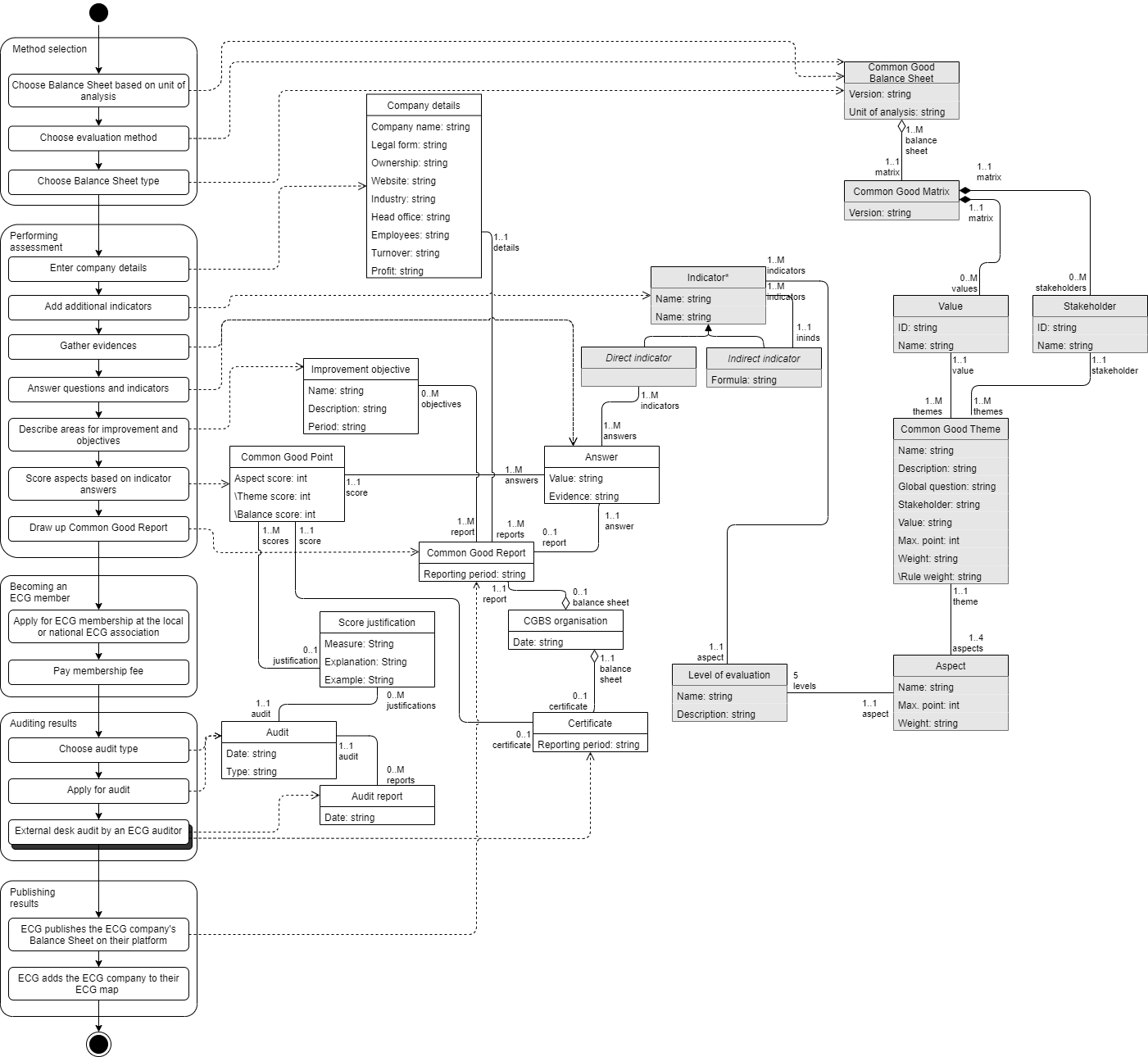}
    \caption{The PDD of the CGBS method}
    \label{fig:CGBS}
\end{figure}

\subsection{EFQM Model}
\begin{figure}[H]
    \centering
    \includegraphics[width=\textwidth,height=0.7\textheight]{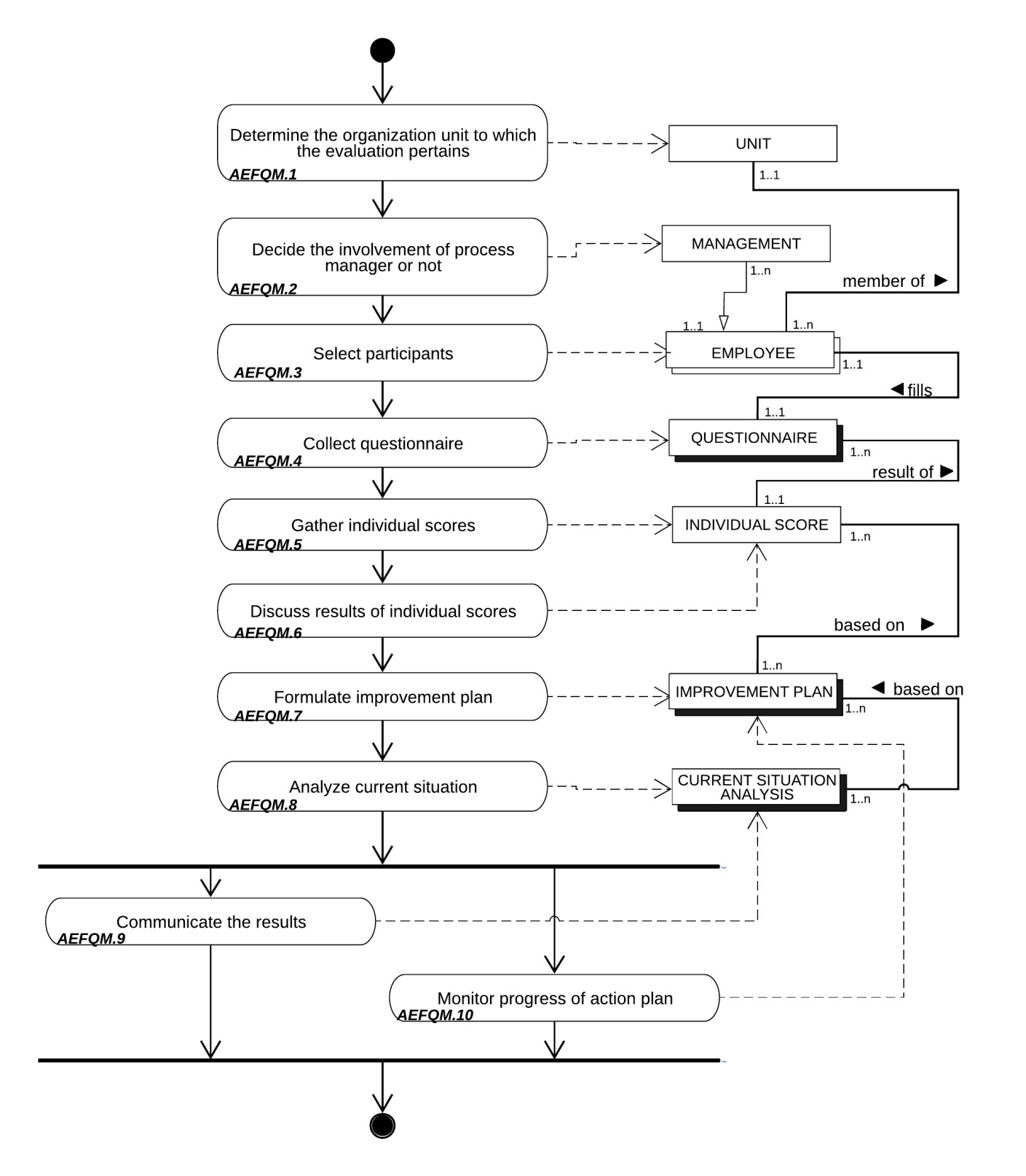}
    \caption{The PDD of the EFQM method}
    \label{fig:EFQM}
\end{figure}

The European Foundation for Quality Management (EFQM) Model, is a self-assessment framework for measuring the strengths and areas for improvement of an organisation across all of its activities \cite{nabitz2000efqm}. Figure~\ref{fig:EFQM} shows the diagram of the self-assessment process and the corresponding deliverables.
 
\newpage

\subsection{Ecovadis}
\begin{figure}[H]
    \centering
    \includegraphics[width=0.8\textwidth]{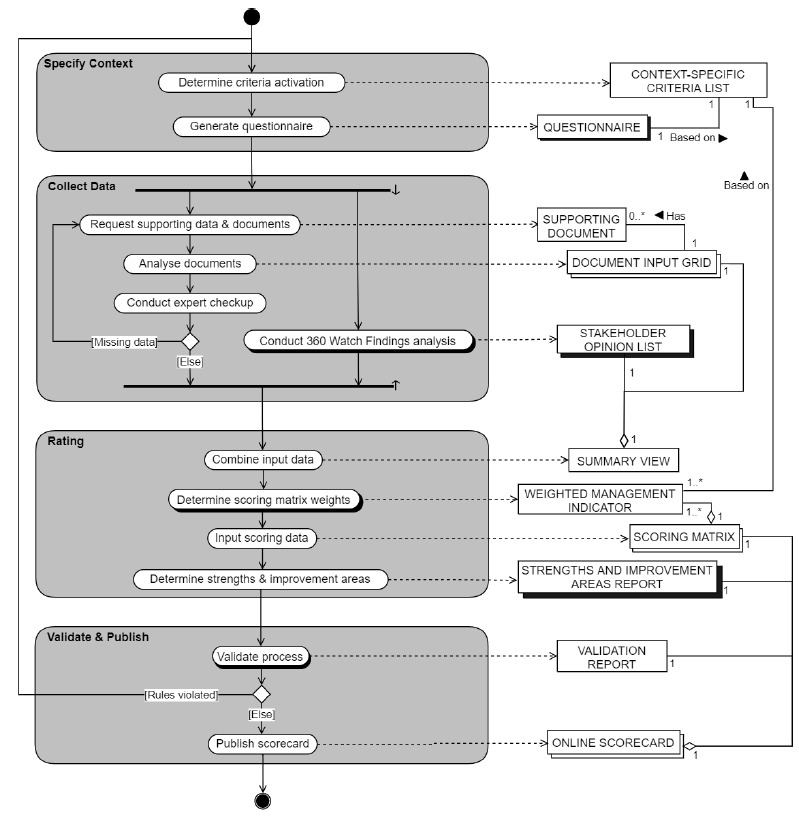}
    \caption{The PDD of the Ecovadis method}
    \label{fig:ecovadis}
\end{figure}

Ecovadis was established in 2007. The company offers an environmental sustainability ratings platform to assess corporate social responsibility and sustainable procurement \cite{ecovadis}. Figure~\ref{fig:ecovadis} depicts the PDD of the sustainability assessment method offered by Ecovadis.

\newpage

\subsection{Fair Trade Software Foundation (FTSF) certification}
Fair Trade is a social movement that aims to help producers in developing countries through partnerships promoting and selling their products. Fair Trade Software extends this concept into the software industry, whilst carefully adhering to the ten commonly accepted Fair Trade Principles established by the World Fair Trade Organization. A preliminary version of the FTSF certification method, that was designed in collaboration with FTSF stakeholders, is depicted in the PDD. The method has not yet been implemented.

\begin{figure}[H]
    \centering
    \includegraphics[width=0.85\textwidth]{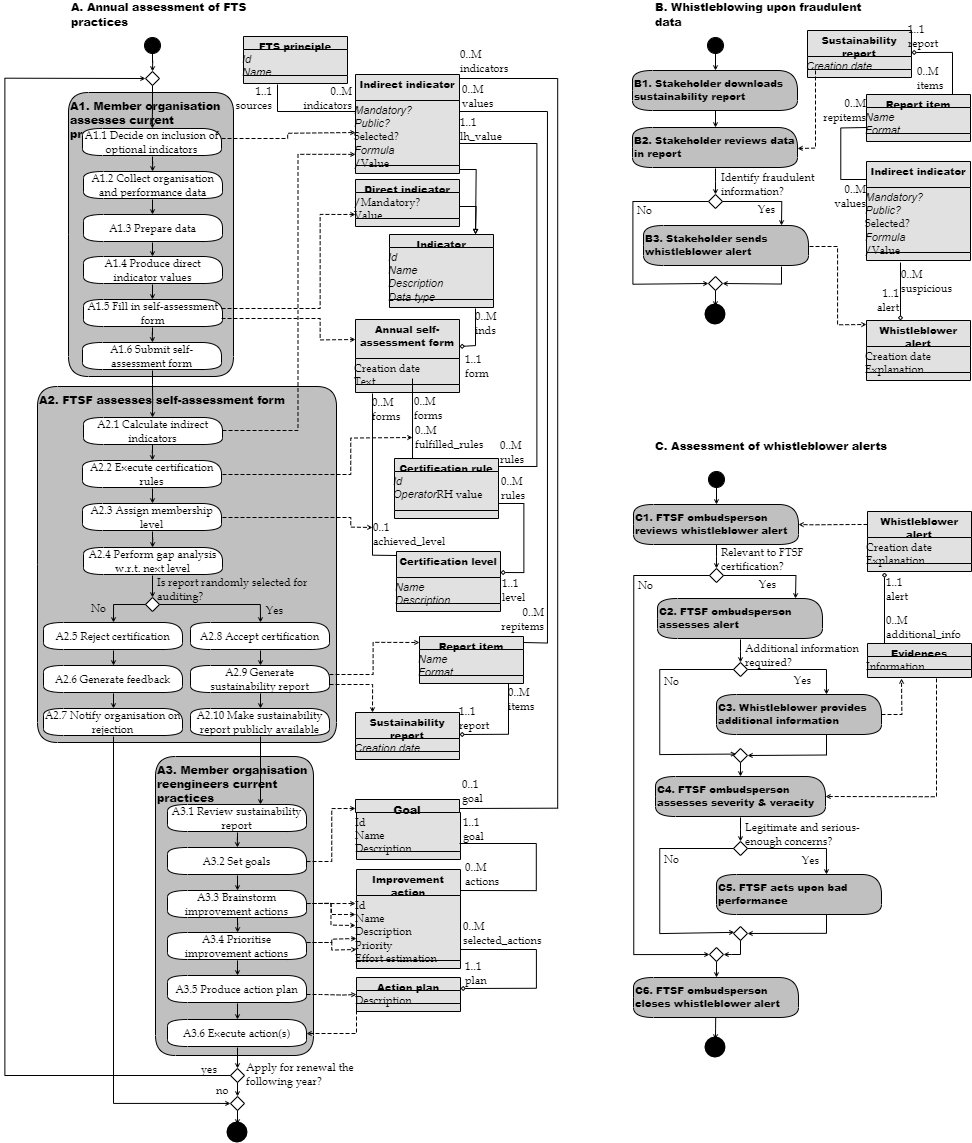}
    \caption{The PDD of the FTSF method}
    \label{fig:FTSF}
\end{figure}

\newpage

\subsection{Greenhouse Gas Protocol}
The Greenhouse Gas Protocol (GHG) Protocol Corporate Accounting and Reporting Standard provides requirements and guidance for companies and other organisations, such as NGOs, government agencies, and universities, that are preparing a corporate-level GHG emissions inventory \cite{wbcsd2004greenhouse}.

\begin{figure}[H]
    \centering
    \includegraphics[width=\textwidth]{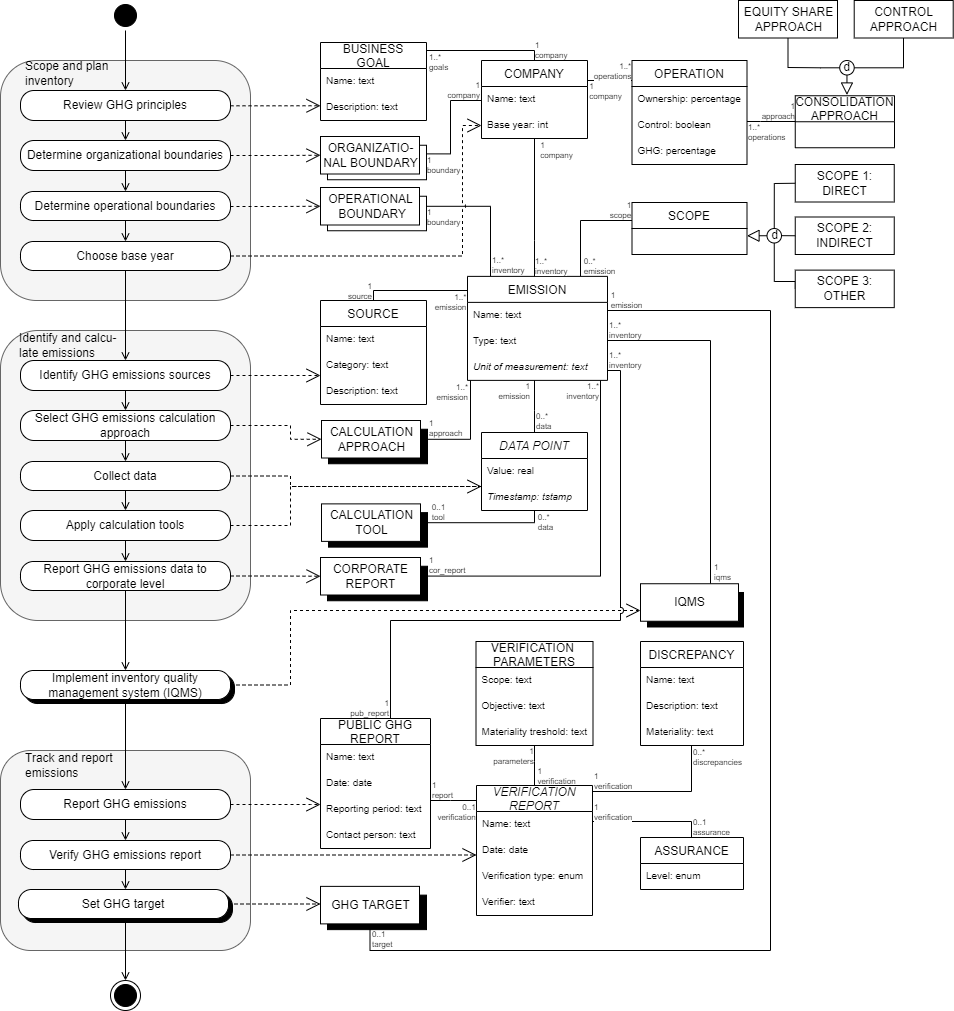}
    \caption{The PDD of the GHG protocol method}
    \label{fig:GHG}
\end{figure}

\newpage

\subsection{Data Centre Assessment}
The Data Centre Assessment by Green IT Switzerland is a questionnaire, that assesses the maturity of data centres with regard to Green IT best practices. The questionnaire is made up of a hierarchical set of pages. Each page contains information and / or questions. Pages can be filled out in any order. The results of the assessment are available at any time, even if not all questions are answered yet.

\begin{figure}[H]
    \centering
    \includegraphics[width=\textwidth]{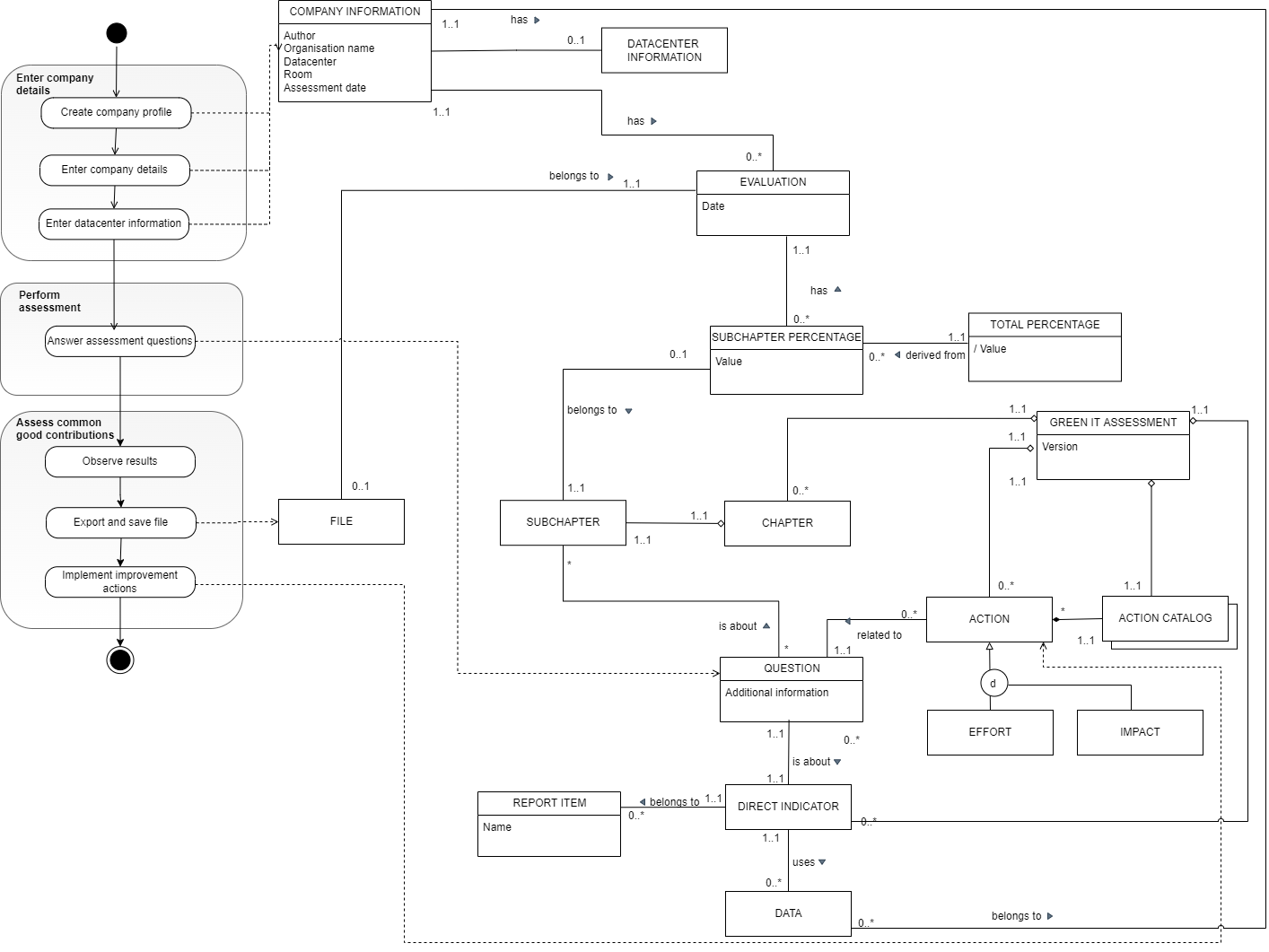}
    \caption{The PDD of the Green IT Assessment method}
    \label{fig:GIT}
\end{figure}

\newpage

\subsection{GRI Standards}

The GRI Standards help organisations understand their outward impacts on the economy, environment, and society, including those on human rights \cite{GRI_Standards}. This increases accountability and enhances transparency on their contribution to sustainable development. The GRI Standards are a modular system comprised of three series of Standards to be used together: Universal Standards, Sector Standards, and Topic Standards. Organisations can either use the GRI Standards to prepare a sustainability report in accordance with the Standards or use selected Standards, or parts of their content, to report information for specific users or purposes, such as reporting their climate change impacts for their investors and consumers. The PDD shows an overview of constructing a sustainability report according to the GRI reporting guidelines.

\begin{figure}[H]
    \centering
    \includegraphics[width=\textwidth, height=0.7\textheight]{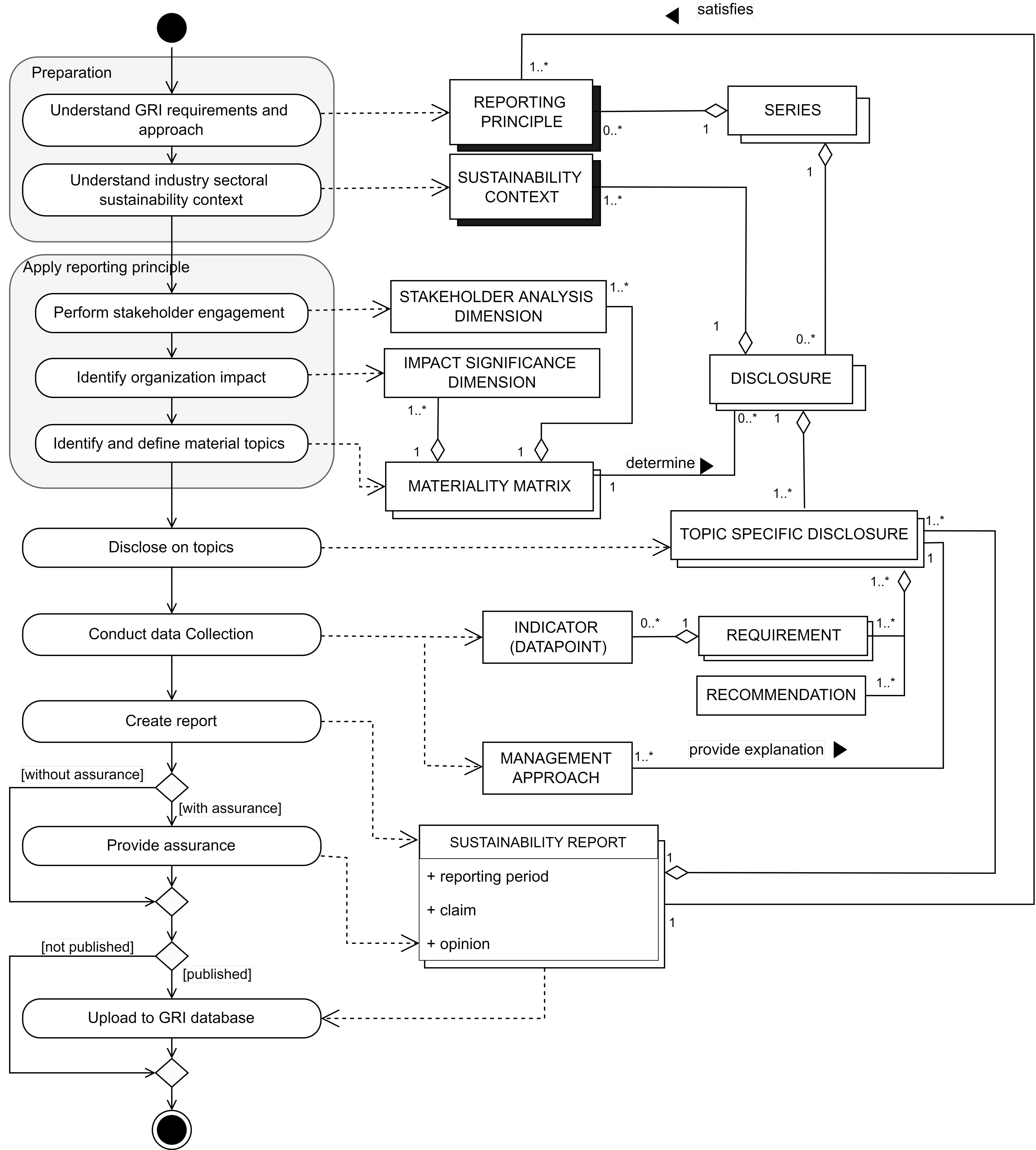}
    \caption{The PDD of the GRI Standards method}
    \label{fig:GRI}
\end{figure}

\subsection{ISO14001}
ISO 14001:2015 sets out the criteria for an environmental management system and can be certified to \cite{ISO14001}. Using ISO 14001:2015 can provide assurance to company management and employees as well as external stakeholders that environmental impact is being measured and improved.

\begin{figure}[H]
    \centering
    \includegraphics[width=\textwidth,height=0.8\textheight]{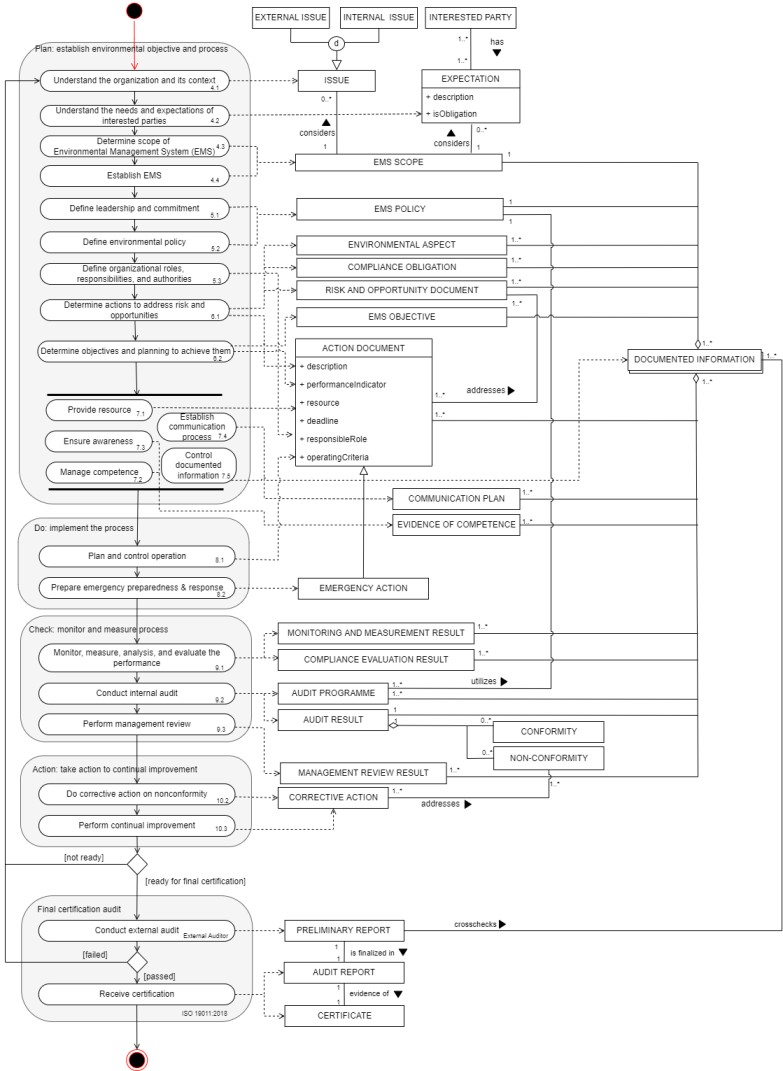}
    \caption{The PDD of the ISO14001 method}
    \label{fig:ISO14001}
\end{figure}

\subsection{ISO26000}
ISO 26000:2010 is intended to provide organisations with guidance concerning social responsibility and can be used as part of public policy activities \cite{moratis2017iso}. ISO 26000:2010 is not a management system standard. It is not intended or appropriate for certification purposes or regulatory or contractual use. As ISO 26000:2010 does not contain requirements, any such certification would not be a demonstration of conformity with ISO 26000:2010.

\begin{figure}[H]
    \centering
    \includegraphics[width=0.8\textwidth,height=0.8\textheight]{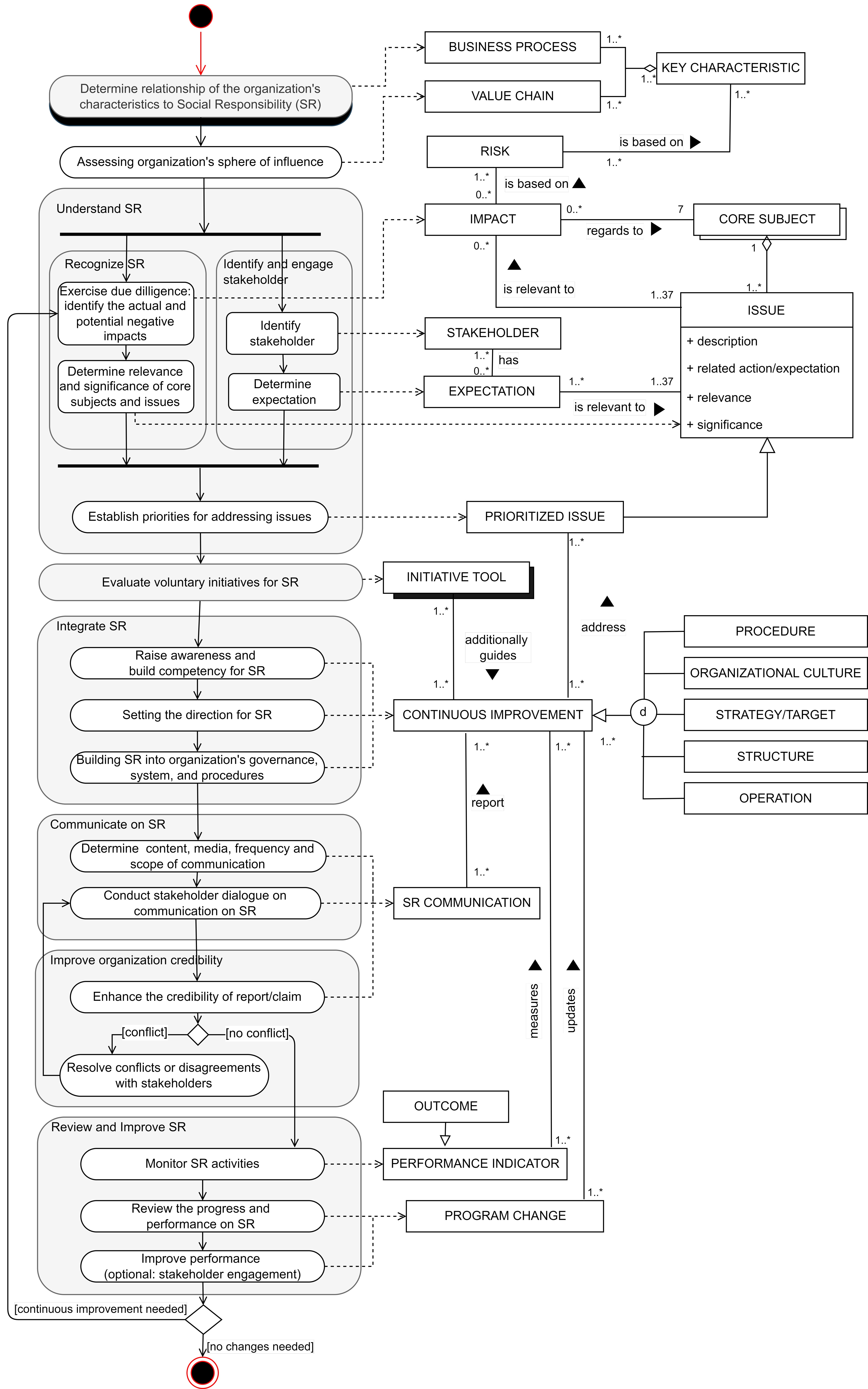}
    \caption{The PDD of the ISO26000 method}
    \label{fig:ISO26000}
\end{figure}

\subsection{S-CORE}
Sustainability – Competency, Opportunity, Reporting and Evaluation (S-CORE) is a web-based tool used to assess where and how sustainability lies within an organisation, while also providing insight into opportunities based on identified sustainability goals. S-CORE is not limited to a certain business type or size, and includes almost 100 practices across various sectors that span all levels and stages of sustainability implementation \cite{hart2016your}.

\begin{figure}[H]
    \centering
    \includegraphics[width=\textwidth]{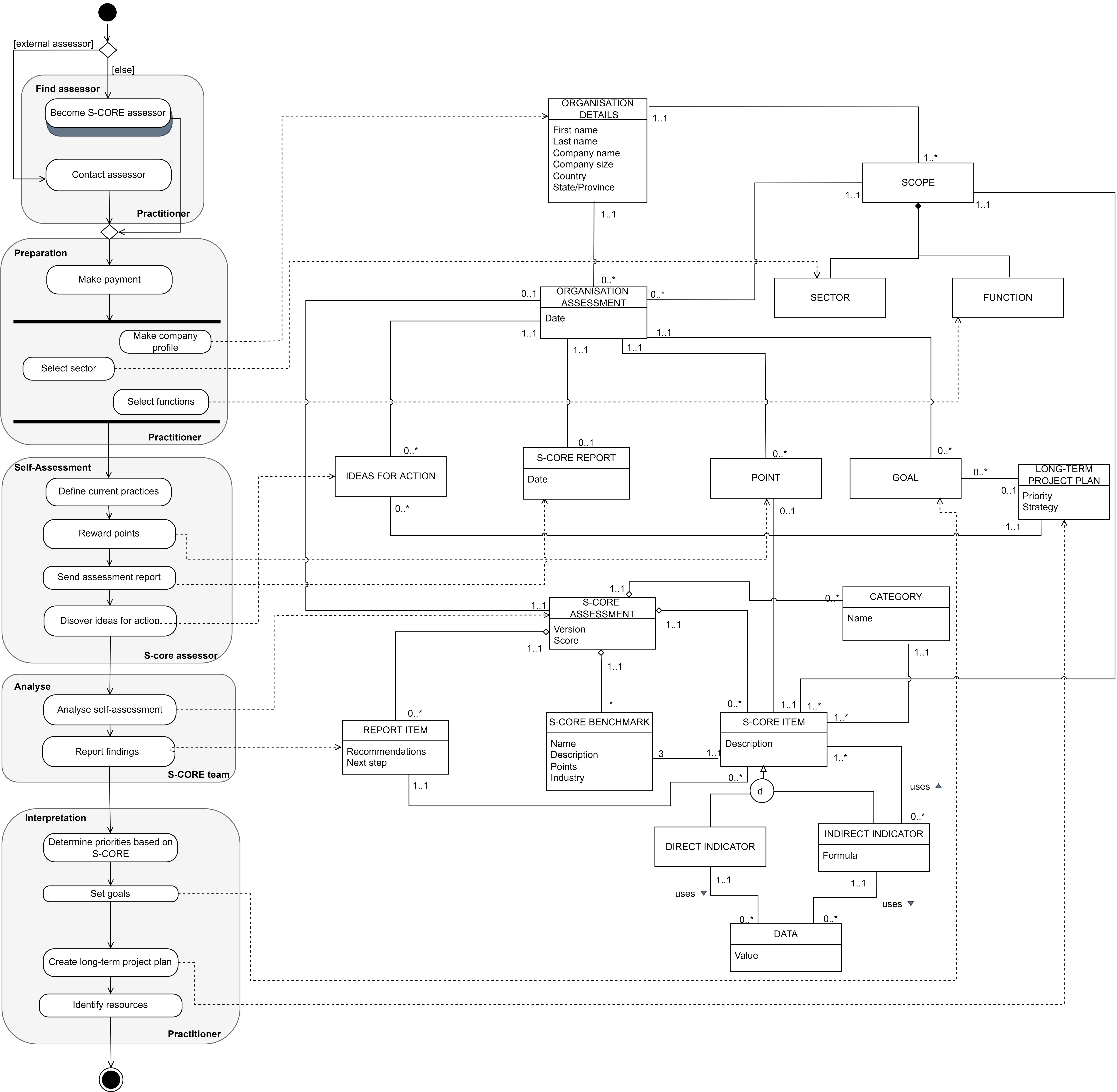}
    \caption{The PDD of the S-CORE method}
    \label{fig:SCORE}
\end{figure}

\newpage

\subsection{NCP}
The Natural Capital Protocol is a decision-making framework that enables organisations to identify, measure and value their direct and indirect impacts and dependencies on natural capital \cite{whitaker2018natural}. It is focused at a business decision-making level and helps organisations to understand the value of their dependence on ecosystem flows, rather than the value of natural capital stocks.

\begin{figure}[H]
    \centering
    \includegraphics[height=0.8\textheight]{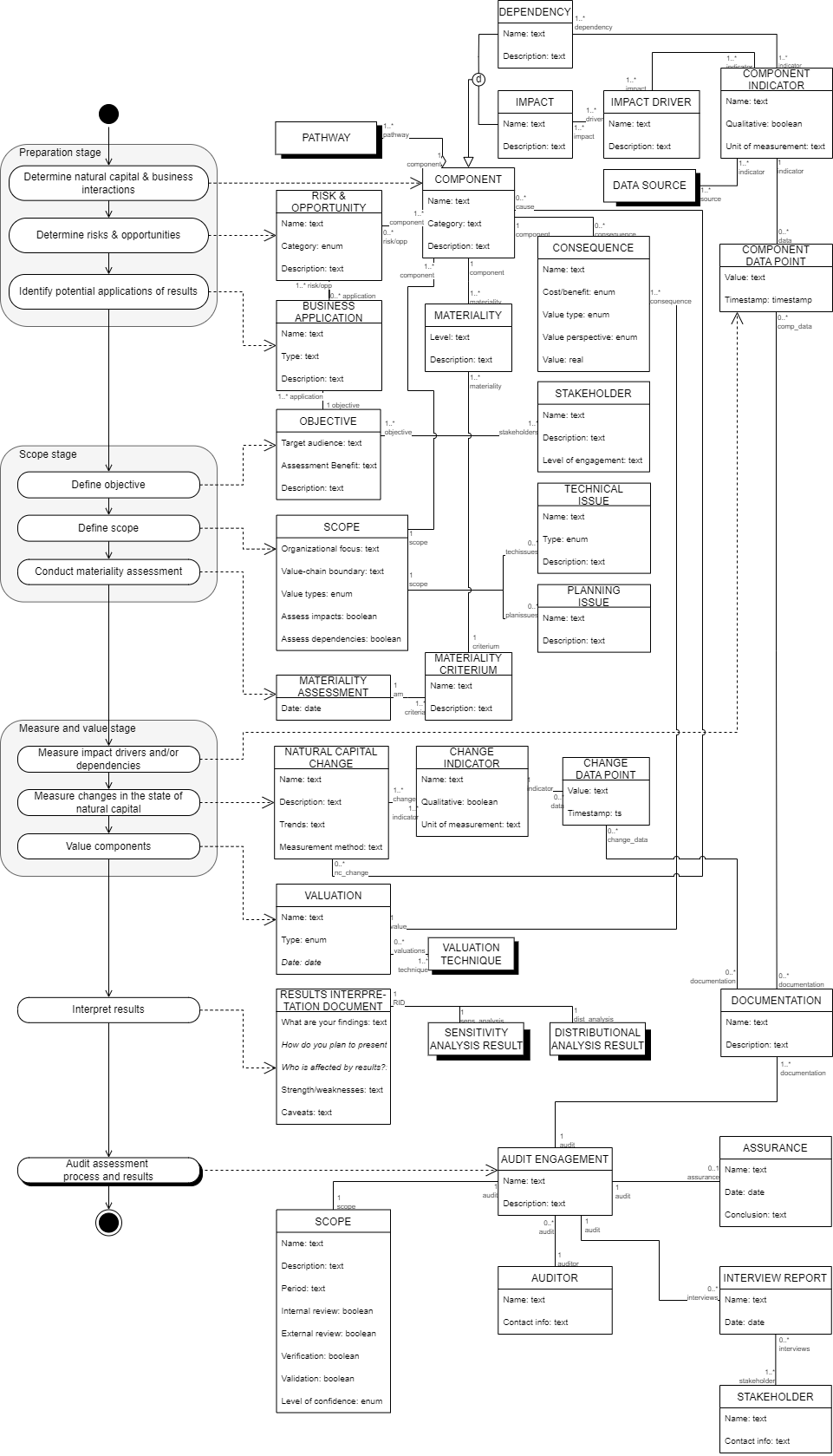}
    \caption{The PDD of the NCP method}
    \label{fig:NCP}
\end{figure}

\newpage

\subsection{Sustainable Development Goals Compass}
The Sustainable Development Goals (SDG) Compass was developed to meet the uncertainty about what actions an organisation can and should take in order to contribute to the goals, the SDG Compass is a guide that companies can use to align their strategies with the relevant SDGs, and measure and manage their impacts \cite{briones2021sdg}.

\begin{figure}[H]
    \centering
    \includegraphics[width=\textwidth,height=0.8\textheight]{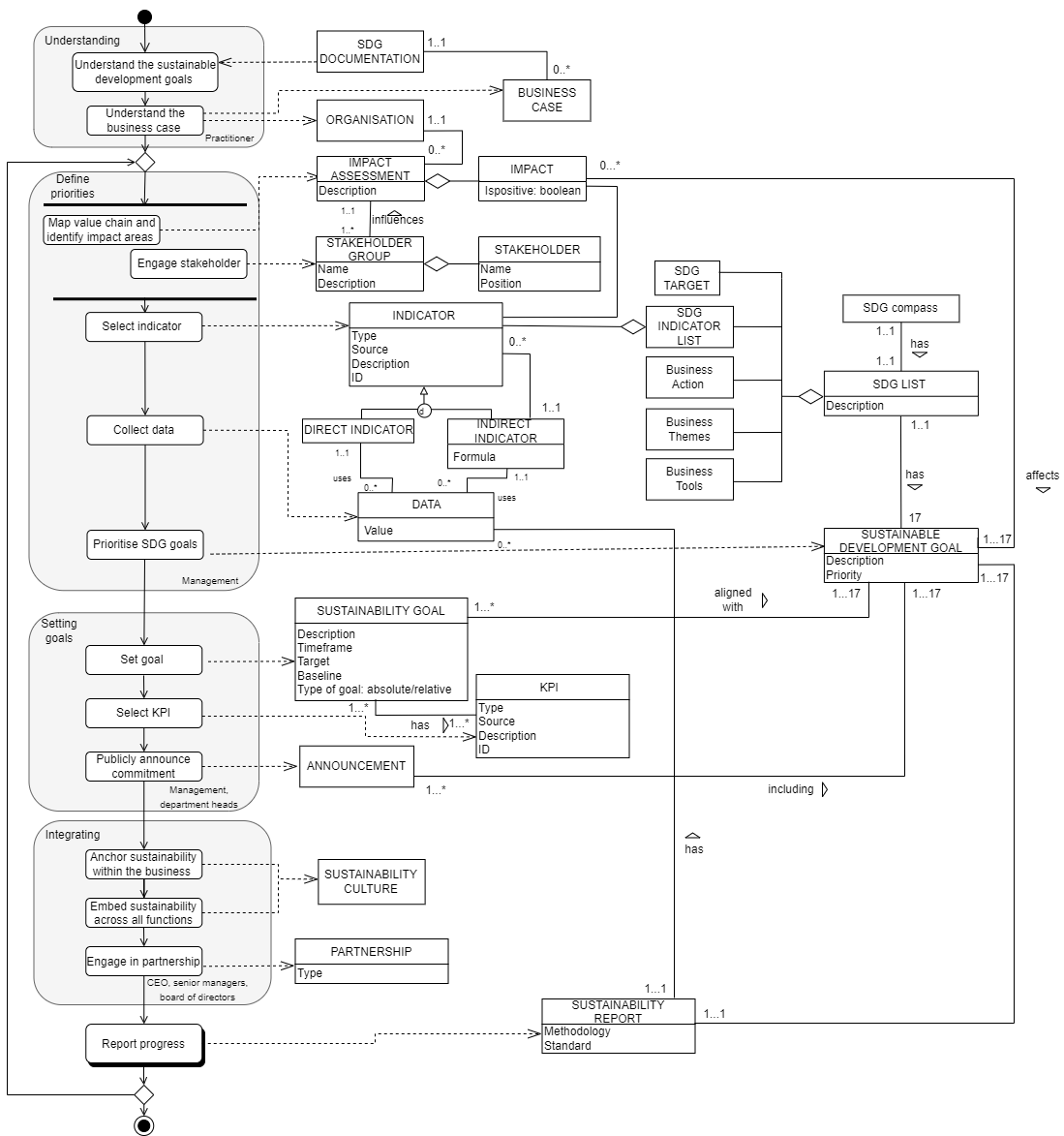}
    \caption{The PDD of the Sustainable Development Goals Compass method}
    \label{fig:SDGc}
\end{figure}

\subsection{SMETA}
Sedex Members Ethical Trade Audit (SMETA) is an ethical accounting and audit method which encompasses aspects of responsible business practice. As a multi-stakeholder initiative, SMETA was designed to minimise duplication of effort and provide members and suppliers with an audit format they could easily share \cite{medina2016application}. SMETA reports are published in the SEDEX system, ensuring transparency and efficient information sharing.

SMETA audits use the ETI Base Code, founded on the conventions of the International Labor Organization, as well as relevant local laws. SMETA audits can be conducted against two or four auditing pillars. The two pillars mandatory for any SMETA audit are Labor Standards and Health \& Safety. The two additional pillars of a 4-pillar audit are Business Ethics and Environment. They were introduced to further deepen the social responsibility aspect of SMETA audits.

\begin{figure}[H]
    \centering
    \includegraphics[height=0.75\textheight]{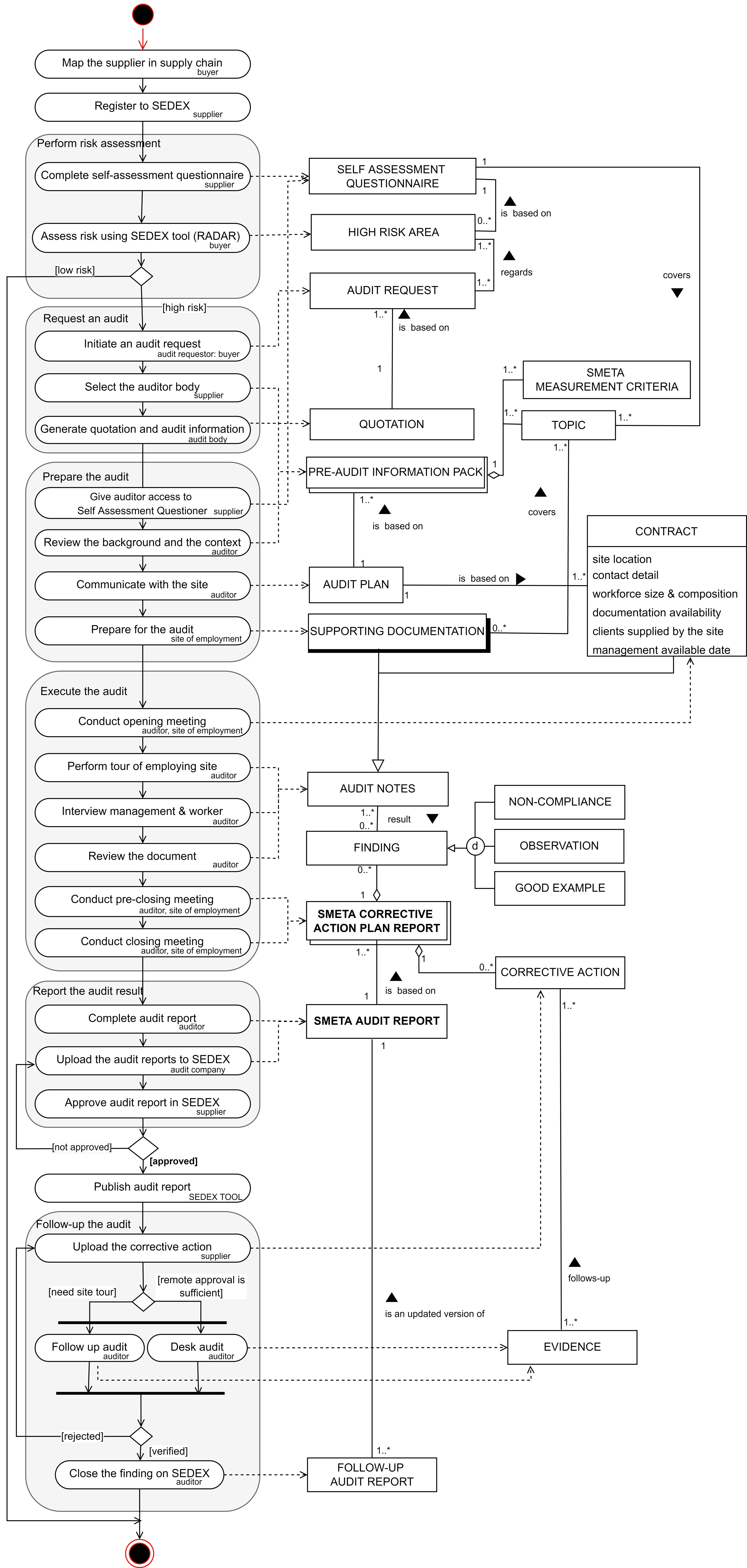}
    \caption{The PDD of the SMETA method}
    \label{fig:SMETA}
\end{figure}

\subsection{STARS}
The Sustainability Tracking, Assessment \& Rating System (STARS) is a transparent, self-reporting framework for colleges and universities to measure their sustainability performance \cite{urbanski2015measuring}. STARS is intended to engage and recognise the full spectrum of higher education institutions, from community colleges to research universities. It encompasses long-term sustainability goals for already high-achieving institutions, as well as entry points of recognition for institutions that are taking first steps toward sustainability. 

\begin{figure}[H]
    \centering
    \includegraphics[width=\textwidth]{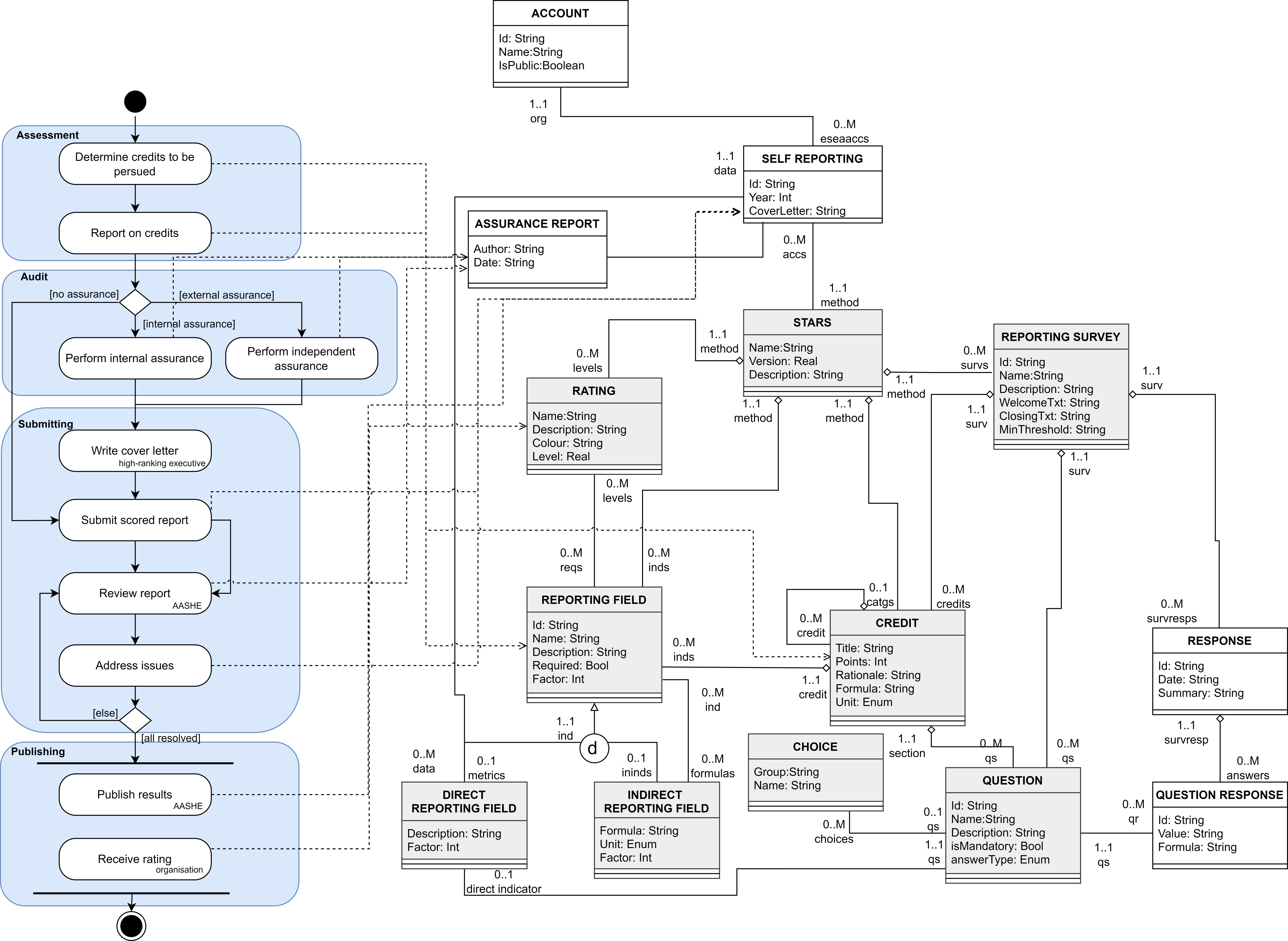}
    \caption{The PDD of the STARS method}
    \label{fig:STARS}
\end{figure}

\newpage

\subsection{UNGC}
\begin{figure}[H]
    \centering
    \includegraphics[width=0.7\textwidth]{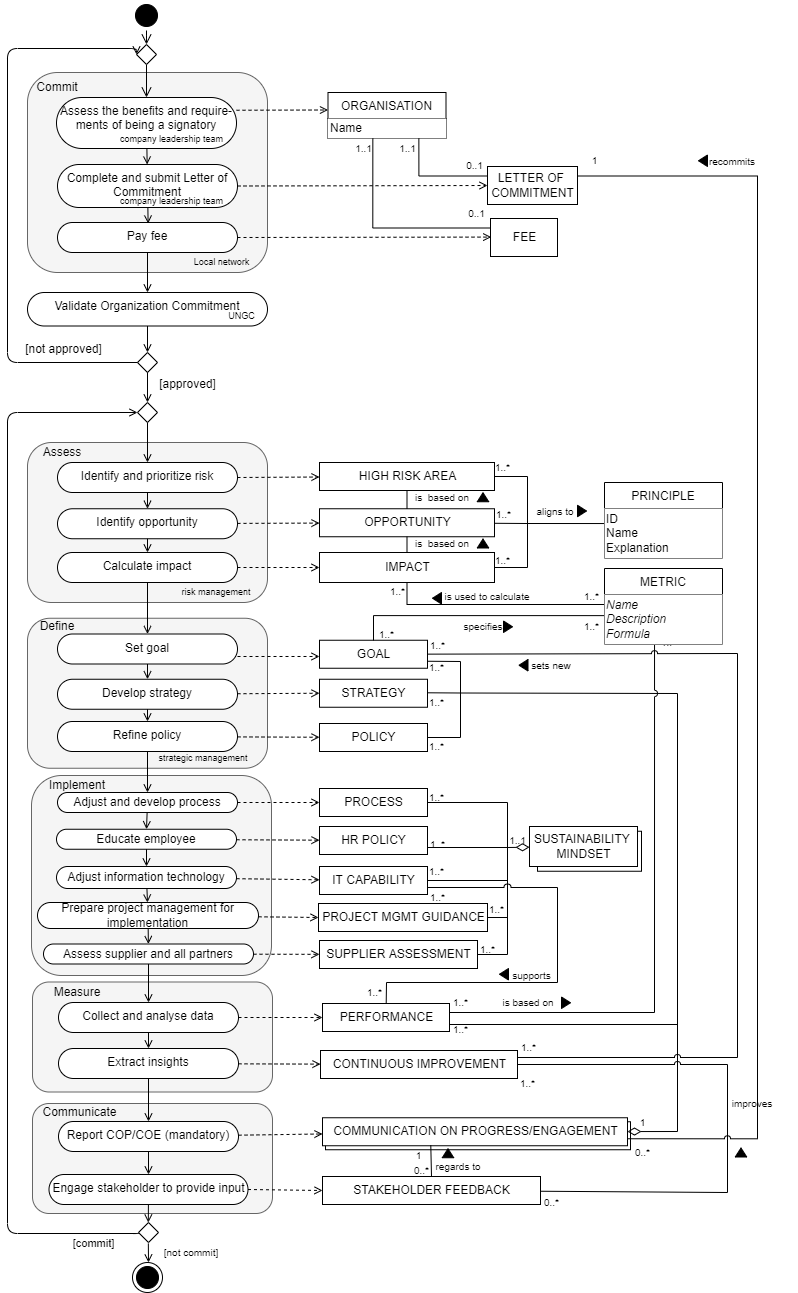}
    \caption{The PDD of the UNGC method}
    \label{fig:UNGC}
\end{figure}

The United Nations Global Compact is a non-binding United Nations pact to encourage businesses and firms worldwide to adopt sustainable and socially responsible policies, and to report on their implementation \cite{williams2004global}. The PDD of the ESEA method that ensures that companies adhere to the UNGC principles is shown in Figure\ref{fig:UNGC}

\subsection{UniSAF}
The University Sustainability Assessment Framework (UniSAF) is developed by rootAbility, a non-profit social business that promotes
sustainability projects and initiatives in higher education. The method provides a set of indicators and methodology that can be used to gather and analyse data on the sustainability performance of university \cite{uniSAF}. 

\begin{figure}[H]
    \centering
    \includegraphics[width=\textwidth,height=0.8\textheight]{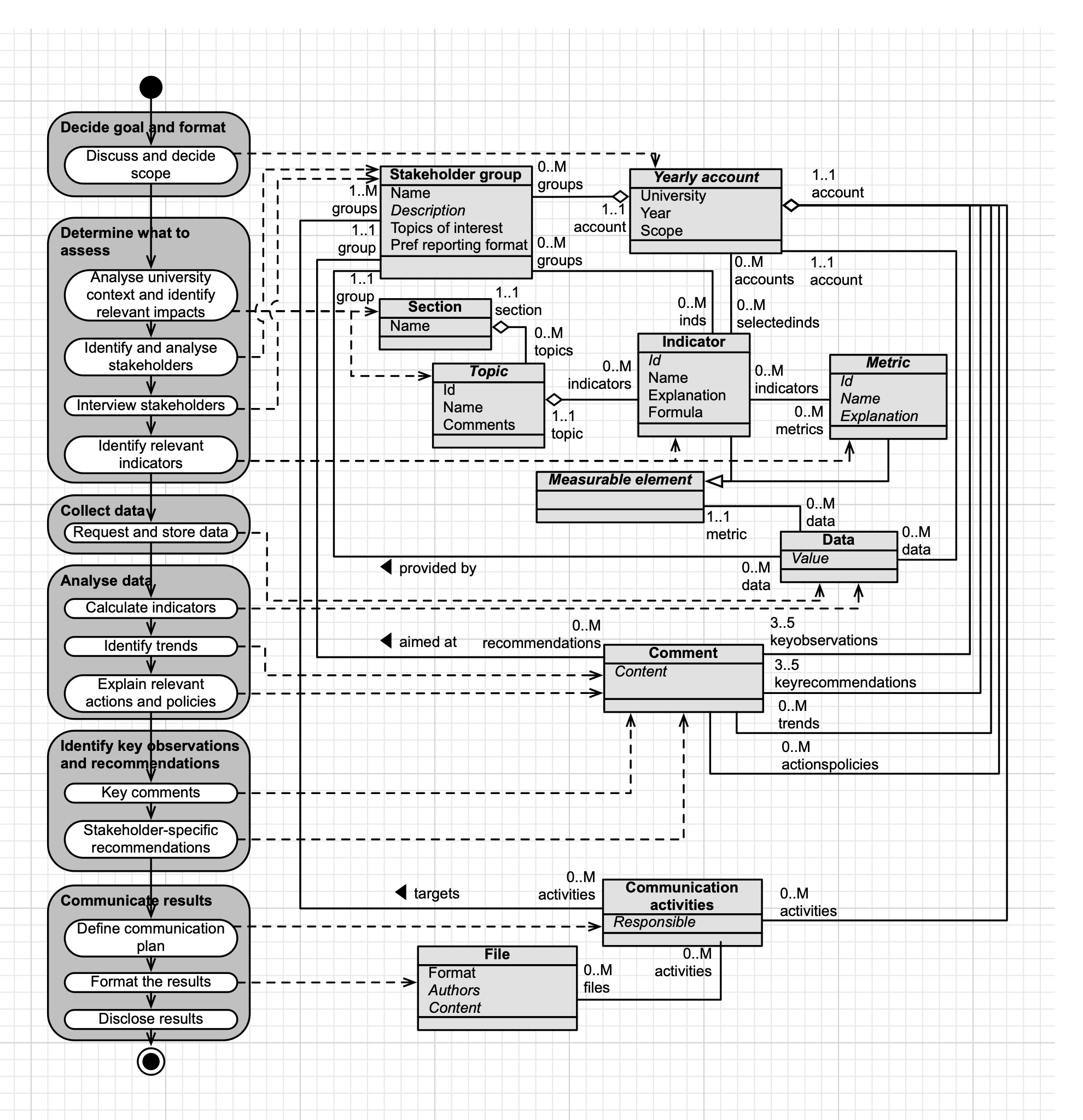}
    \caption{The PDD of the UniSAF method}
    \label{fig:UniSAF}
\end{figure}

\subsection{XES Social Balance}
The XES Social Balance is used by the Catalan Network of Solidarity Economy (XES) to assess the performance of their members. The method has two variants: Basic Social Balance, Complete Social Balance. The complete one has more direct indicators (questions) and indirect indicators (indicators), and it also deploys multi-respondent surveys to several stakeholder groups (depending on the context of the organisation): workers/members, clients/suppliers, volunteers \cite{crusellas2019balance}.

\begin{figure}[H]
    \centering
    \includegraphics[width=\textwidth]{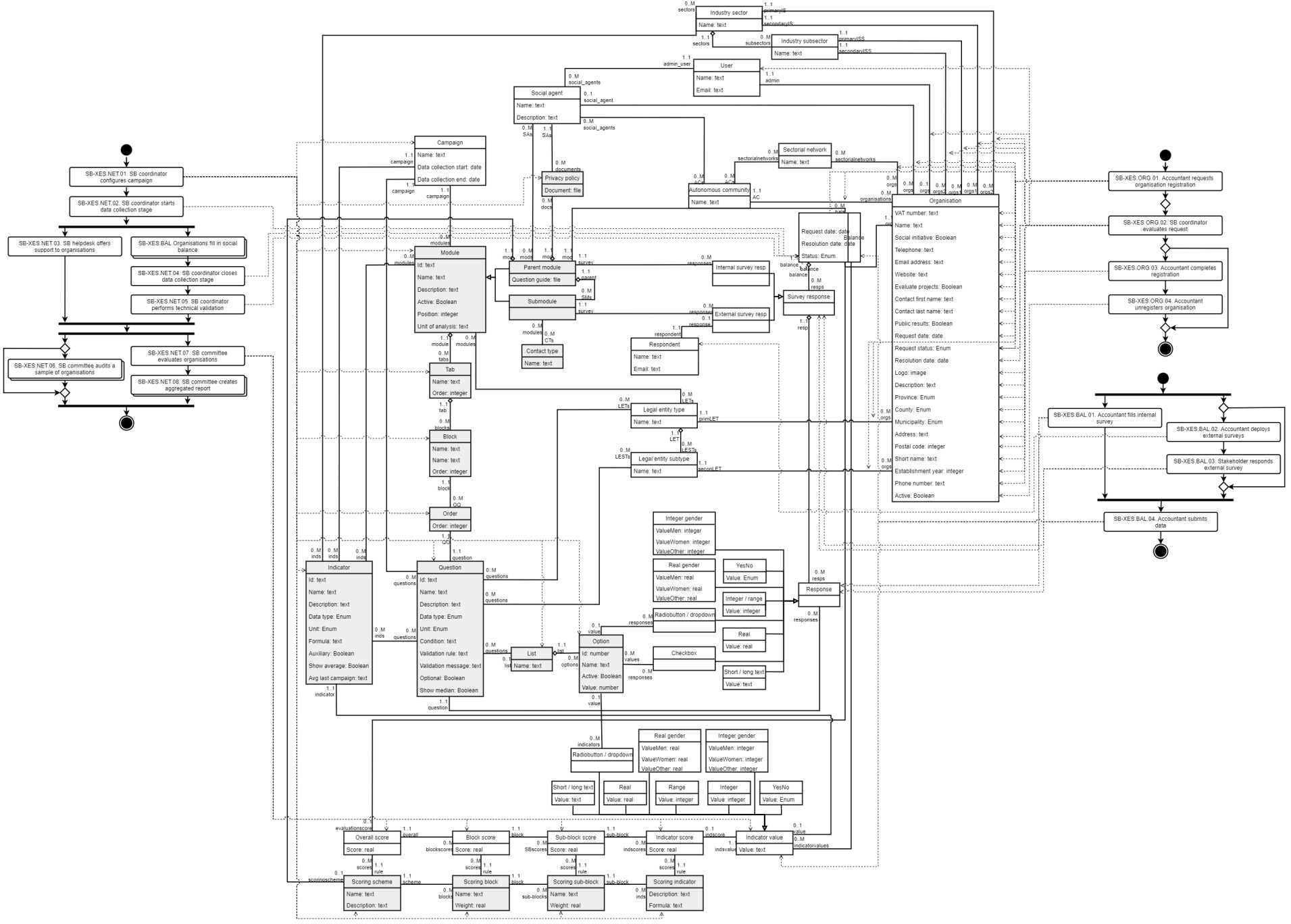}
    \caption{The PDD of the XES method}
    \label{fig:XES}
\end{figure}

\section{Discussion of results}
\label{sec:discussion}
\begin{figure}[H]
    \centering
    \includegraphics[width=\textwidth]{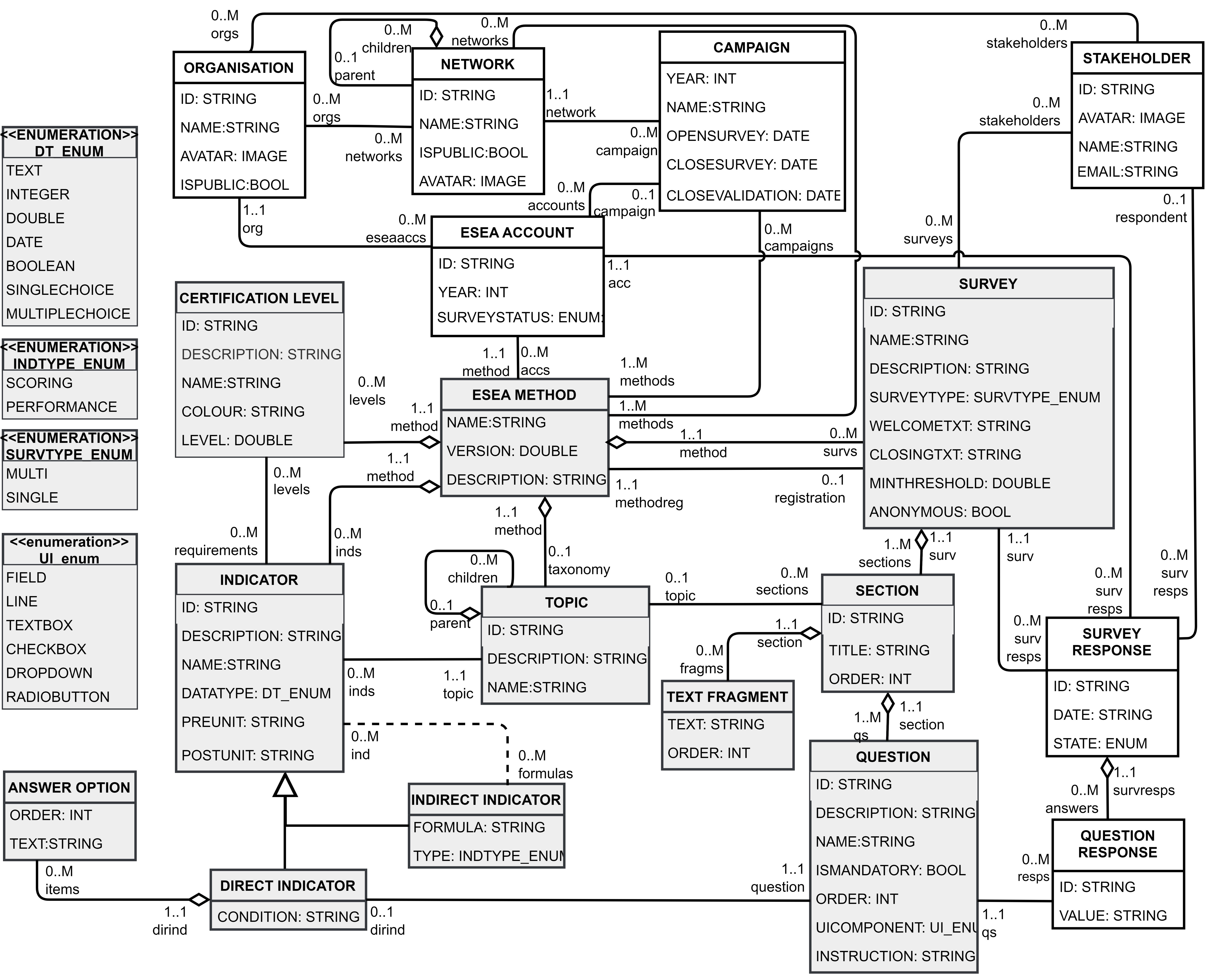}
    \caption{The openESEA metamodel, further specified in \cite{TR_modelling_language}}
    \label{fig:metamodel}
\end{figure}

The openESEA metamodel is created based on the PDDs of ESEA methods. The PDDs are compared using the method comparison approach. This approach yields the metamodel depicted in Figure~\ref{fig:metamodel}. An earlier version of the metamodel is presented in~\cite{espana2019model}. Based on the current version we have engineered an Xtext grammar. Using the Xtext grammar method engineers can create models of ESEA methods. Such models can be interpreted by our open-source model-driven tool, likewise called openESEA\footnote{\url{https://github.com/sergioespana/openESEA}}. To test the comprehensibility and usability of the openESEA metamodel and grammar we ran an experiment. The results of the experiment are presented in \cite{BIR}.

\section{Conclusion}
\label{sec:conclusion}
Over the course of the years our collection of ESEA method diagrams has grown rapidly. The models have helped us better understand ESEA methods and they have laid foundations for further analysis and software design and development. The domain analysis presented in this report has enabled several project already. We have created the first version of model-driven ESEA method interpreter \cite{espana2019model}, an extension of the tool that generates infographics automatically \cite{espana2022domain}, the investigation of ESEA method selection criteria \cite{ECGIC}, the engineering of the openESEA DSL \cite{BIR}. In the future we plan to publish information on ESEA methods, as well as the PDDs in an online repository. 

\section{Acknowledgements}
\label{sec:acknowledgements}
We thank all the students of Utrecht University's Business Informatics master's programme who modelled, validated, or improved PDDs as part of their graduation projects. Moreover, we express our gratitude to all the ESEA method experts who helped modelling and validating the diagrams. Without the contributions of all collaborations we would not have been able to present such a detailed collection of ESEA method models. 

\bibliographystyle{unsrtnat}
\bibliography{references} 

\appendix
\section*{Appendix A: List of analysed ESEA method}

\begin{table}[H]
    \caption{The list of all analysed ESEA methods}
    \centering
    \scalebox{0.7}{
    \begin{tabular}{|p{6cm}|p{2cm}|p{6cm}|p{8cm}|}
    \hline
      \textbf{Method}  & \textbf{Release year} & \textbf{Organisation} & \textbf{URL}  \\ \hline
      
       AA1000AS  & 1999 & AccountAbility & \url{https://www.accountability.org/standards/aa1000-assurance-standard/} \\ \hline
       
       B Impact Assessment & 2007 & B Lab & \url{https://bimpactassessment.net/} \\ \hline
       
       CDP  & 2016 & CDP & \url{https://www.cdp.net/en/companies} \\ \hline
       
       Common Good Balance Sheet  & 2010 & Economy for the Common Good & \url{https://www.ecogood.org/apply-ecg/companies/\#balance-sheet-resources} \\ \hline
       
       EcoVadis  & 2007 & EcoVadis & \url{https://ecovadis.com/} \\ \hline
       
       EFQM Model  & 1992 & European Foundation for Quality Management (EFQM) & \url{https://www.efqm.org/efqm-model/} \\ \hline
       
       FTSF Certification  & 2010 & Fair Trade Software Foundation & \url{https://ftsf.eu/about-us} \\ \hline

        Greenhouse Gas Protocol & 2001 & Greenhouse Gas Protocol & \url{https://ghgprotocol.org/standards} \\ \hline
        
        Data Center Assessment & 2017 & Green IT Switzerland & \url{https://greenit-switzerland.ch/app/dca/en/dc-assessment} \\ \hline
        
        GRI Standards & 2000 & Global Reporting Initiative  & \url{https://www.globalreporting.org/how-to-use-the-gri-standards/} \\ \hline
        
        ISO14001 & 2004 & International Organization of Strandardization & \url{https://www.iso.org/iso-14001-environmental-management.html} \\ \hline
        
        ISO26000 & 2010 & International Organization of Strandardization & \url{https://www.iso.org/iso-26000-social-responsibility.html} \\ \hline 
        
        Natural Capital Protocol & 2016 & Capitals Coalition & \url{https://capitalscoalition.org/capitals-approach/natural-capital-protocol/?fwp_filter_tabs=training_material} \\ \hline  
        
        S-CORE & 2005 & International Society of Sustainability Professionals & \url{https://sustainablemeasures.com/} \\ \hline   
        
        Sustainable Development Goals Compass & 2015 & GRI, the UN Global Compact and the World Business Council for Sustainable Development (WBCSD) & \url{https://sdgcompass.org/download-guide/} \\ \hline   
        
        SMETA & 2017 & Sedex & \url{https://www.sedex.com/our-services/smeta-audit/} \\ \hline   
        
        STARS & 2007 & Association for the Advancement of Sustainability in Higher Education (AASHE) & \url{stars.aashe.org} \\ \hline   
        
        UN Global Compact & 2000 & United Nations & \url{https://www.unglobalcompact.org/what-is-gc/mission/principles} \\ \hline   
        
        UniSAF & 2017 & Student Organization for Sustainability International & \url{https://www.greenofficemovement.org/sustainability-assessment/} \\ \hline   
        
        WFTO certification & 2013 & World Fair Trade Organization  & \url{https://wfto.com/what-we-do\#our-guarantee-system} \\ \hline  
        
        XES Social Balance & - & Catalan Network of Solidarity Economy & \url{https://xes.cat/} \\ \hline  
    \end{tabular}}
    \label{tab:Method_List}
\end{table}

\end{document}